\documentclass[prl,twocolumn,superscriptaddress]{revtex4-1}
\usepackage{epsfig}
\usepackage{bm}
\usepackage{amsmath}
\usepackage{amssymb,color}
\usepackage[Symbol]{upgreek}

\usepackage[unicode=true,bookmarks=true,bookmarksnumbered=false,bookmarksopen=false,breaklinks=false,pdfborder={0 0 1},backref=false,colorlinks=true]{hyperref}

\hypersetup{linkcolor=magenta,urlcolor=blue,citecolor=blue,pdfstartview={FitH},hyperfootnotes=false,unicode=true}

\def\be{\begin{equation}}
\def\ee{\end{equation}}
\def\bea{\begin{eqnarray}}
\def\eea{\end{eqnarray}}
\def\nn{\nonumber}

\begin{document}

\title{Observation of a dynamical sliding phase superfluid with P-band bosons}
\author{Linxiao Niu}
\author{Shengjie Jin} 
\author{Xuzong Chen} 
\affiliation{School of Electronics Engineering and Computer Science, Peking University, Beijing 100871, China}
\author{Xiaopeng Li}
\email{xiaopeng\_li@fudan.edu.cn}
\affiliation{State Key Laboratory of Surface Physics, Institute of Nanoelectronics and Quantum Computing, and Department of Physics, Fudan University, Shanghai 200433, China}
\affiliation{Collaborative Innovation Center of Advanced Microstructures, Nanjing 210093, China} 
\author{Xiaoji Zhou} 
\email{xjzhou@pku.edu.cn}
\affiliation{School of Electronics Engineering and Computer Science, Peking University, Beijing 100871, China}

\begin{abstract}


Sliding phases have been long sought-after in the context of coupled XY-models, of relevance to various many-body systems such as layered superconductors, free-standing liquid-crystal films, and cationic lipid-DNA complexes. Here we report an observation of a dynamical sliding-phase superfluid that emerges in a nonequilibrium setting from the quantum dynamics of a three-dimensional ultracold atomic gas loaded into the  P-band of a one-dimensional optical lattice.  A shortcut loading method is used to transfer atoms into the P-band at zero quasi-momentum within a very short time duration.  The system can be viewed as a series of ``pancake"-shaped atomic samples. For this far-out-of-equilibrium system, we find an intermediate time window with lifetime around tens of milliseconds, where the atomic ensemble exhibits robust superfluid phase coherence in the pancake directions, but no coherence in the lattice direction, which implies a dynamical sliding-phase superfluid. The emergence of the sliding phase is attributed to a mechanism of cross-dimensional energy transfer in our proposed phenomenological theory, which is consistent with experimental measurements. This experiment potentially opens up a novel venue to search for exotic dynamical phases by creating high-band excitations in optical lattices.

\end{abstract}
 \maketitle

A sliding phase~\cite{o1999sliding} mechanism has been proposed in the study of weakly coupled stacks of XY models~\cite{granato1986critical,choi1985phase}, which  was introduced to characterize intricate phase transitions in a broad range of many-body systems such as layered superconductors~\cite{feigel1990pinning,klemm1975theory}, free-standing liquid-crystal films~\cite{stoebe1994unusual,cheng1987electron}, and even biological molecules~\cite{o1998sliding,golubovic1998fluctuations}. In the sliding phase, 
{the system behaves like a stack of decoupled superfluid-layers in spite of the physical inter-layer Josephson coupling being finite.} 
With field theory analysis, it has been shown that the sliding phase typically appears under extreme conditions for thermal equilibrium systems~\cite{o1999sliding} or quantum ground states~\cite{Emery2000,Kane2001,Vishwanath2001,Sondhi2001,Kane2002,Liarxiv2017}, causing a grievous challenge in experimental implementation.

\begin{figure}[htp]
\begin{center}
\includegraphics[trim = 0mm 0mm 0mm 0mm, clip=true, width= .83\linewidth]{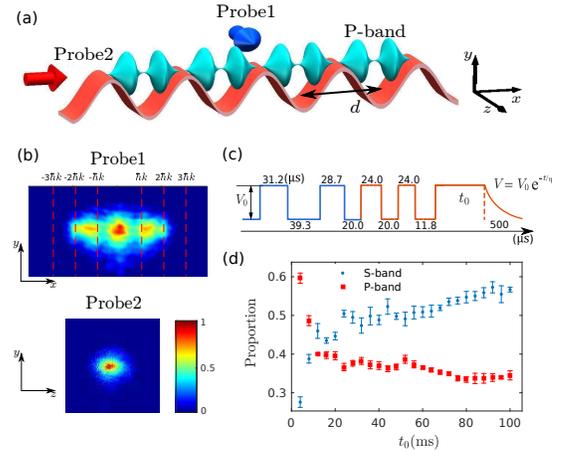}
\end{center}
\caption{(Color online)  
(a) Experimental configuration for a 1d P-orbital lattice, where  atoms form discrete pancakes. 
(b) The system is probed in two ways. Probe1: probe with a laser beam along $\hat{z}$-direction after the band mapping. Probe2: with the probe beam along the $\hat{x}$-direction, the image is taken by switching off the potential abruptly within $30$~ns. Atoms are loaded to the zero quasi-momentum state of the P-band  through a designed pulse sequence, with an example  shown in (c) for lattice-depth $V_{0}=5E_r$ ($E_r$ is one-photon recoil energy). 
We then hold atoms for  time $t_0$, and the absorption images after {time of flight} (TOF) are taken in two directions.  (d) The atom proportion in $S$- and $P$-band, errorbars are given by the standard deviation of 5 experiments.
}
\label{fig:f1}
\end{figure}

Recent experimental progress in synthetic quantum systems  has achieved unprecedented approaches to investigate fascinating collective phenomena in controllable quantum dynamics, such as light-induced non-equilibrium superconductivity~\cite{Fausti2011,Mitrano2016}, time-crystals in trapped ions~\cite{li2012space}, correlated quantum kinematics in reduced-dimensional systems~\cite{Randy,Jorg,Cheng}, and many-body localization with cold atoms in artificial light-crystals~\cite{BAA2006,schreiber2015observation,Kondov2015,Bordia2016,Bordia2017,Choi2016}. 
While a complete theoretical framework to describe nonequilibrium phase-transition is still lacking, a formal analogy between temperature and time by comparing partition-function in the thermal-ensemble and unitary-evolution operator in quantum dynamics allows such concepts in statistical physics as many-body phases and condensation, to generalize to the time-domain~\cite{Renbao2012,Heyl2013}.

\begin{figure*}[htp]
\begin{center}
\includegraphics[trim = 0mm 0mm 0mm 0mm, clip=true, width=.6\linewidth]{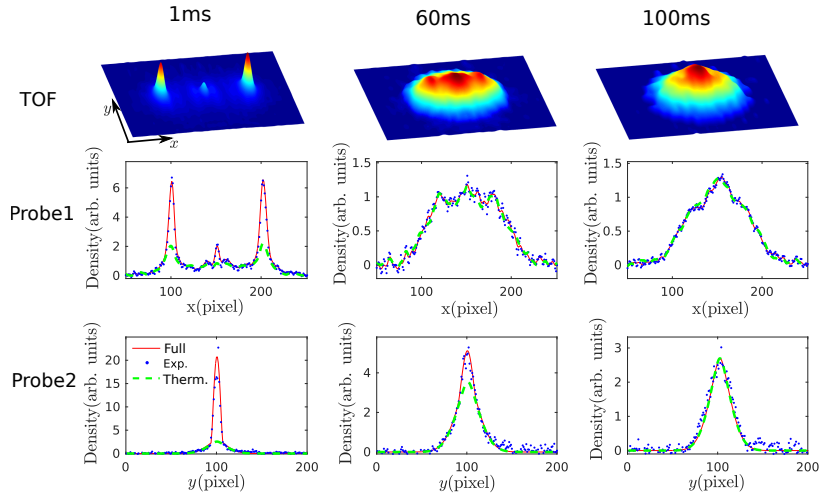}
\end{center}
\caption{ (Color online) 
Momentum distributions measured at three different holding times $t_0$ = 1 ms, 60 ms, 100 ms, along different directions.
The images shown in the first row represent experimental TOF measurements in the $xy$-plane by Probe1 (see main text), 
and the corresponding momentum distribution in the central line along the $\hat{x}$-direction is given in the second row with blue dots. The red solid line gives a full fitting line while the green dashed line gives the distribution of thermal component. 
Atomic distribution in the S-band (P-band) is revealed in the first (second) Brillouin zone. 
The third row shows the atomic distribution along the central line in $\hat{y}$-direction for the
experimental images in the $yz$-plane measured by Probe2.  
Experimental results shown here are taken at temperature $120$~nK and lattice depth $V_0=5E_r$. } 
\label{fig:f2}
\end{figure*}

Here, we report on an observation of a sliding-phase superfluid in
a dynamical system of ultracold atoms loaded into the
P-band of an optical lattice.  Our work goes beyond previous
studies in P-band optical lattices focused on static phases~\cite{Lireview2016,Lewenstein2011,Wirth2011},
by considering nonequilibrium aspects. 
We have a three-dimensional quantum gas confined with a one-dimensional lattice, 
sliced into ``pancakes"  (Fig.~\ref{fig:f1}).
Using an adiabatic short-passage~\cite{load,hu2015long} to
load atoms into the  zero quasi-momentum state in the P-band, the system is driven far-out-of-equilibrium. 
The loading method has been used in our previous experiments to study 
atomic interference between different energy bands~\cite{Ramsey} and long-time evolution~\cite{praWang}. 
Like in previous works the coherent fraction is extracted from interference pattern to explore phase coherence~\cite{prl87,prl92,prl98}.
During the rethermalization process, a metastable region is observed, where the atomic sample shows strong phase-coherence in the
pancake directions, but no coherence in the lattice direction. These
observations imply the first experimental discovery of the sliding-phase superfluid in the time-domain, 
{which is extremely challenging to reach in equilibrium according to the field theoretical analysis~\cite{o1999sliding,Toner1990}.} 
This work may also shed light on  the high-Tc mechanism in light-probed Cuperates~\cite{Fausti2011,Mitrano2016}.


{\it Experimental procedure.---}
The experiment is performed with a BEC of $^{87}$Rb prepared in a hybrid trap 
with the harmonic trapping  frequencies $(\omega _{x},\omega _{y},\omega _{z})=2\pi \times $(28Hz, 55Hz,
60Hz)~\cite{hu2015long}. 
A one-dimensional optical-lattice is produced by a
standing wave  with the lattice constant $d=\pi /k=426$~nm along $x$ axis with $k$ the wave-number. 
As shown in Fig.~\ref{fig:f1}(a), atoms are confined in more than 50 discrete pancakes in the $yz$-plane, and the
sizes of the condensate in the $\hat{y}$ and $\hat{z}$-directions are about $L_y=$15.6~$\upmu$m and $L_z=$14.9~$\upmu$m, respectively. 
{The number of atoms in the trap is about $10^5$.}

A shortcut method with the designed pulse sequences is applied to load atoms into the P-band of the optical-lattice (see Fig.~\ref{fig:f1} and Supplementary Materials). The loading pulse sequence  consists of two sets of pulses whose nodes are shifted in the $\hat{x}$-axis by half of the lattice constant~\cite{load}. We stress here that after loading to the zero quasi-momentum state of the P-band, the quantum system is driven to a far-out-of-equilibrium but at the same time phase-coherent state. 
The short-time collisional dynamics of P-band bosons has previously been observed~\cite{Spielman2006}. We hold the condensate in the P-band for certain amount of time $t_0$ and let the system evolve, then the TOF images are taken after 28 ms of free flight 
in two probe directions---Probe1 and Probe2 to be described below.

For probe from $\hat{z}$-direction (Probe1), the lattice potential is switched off adiabatically, this band mapping (BM) procedure enables measurements of the atomic population in each band.
From such images at different time $t_0$, 
we can quantitatively determine the time-dependent proportion of atoms between S-band and P-band, as shown in Fig.~\ref{fig:f1}(d).  
We investigate the dynamical phase during the decay process from the excited P-band to the S-band.
The probe from $\hat{x}$-direction is performed by an abrupt non-adiabatic switch-off (NAS) of the lattice as shown in Fig.~\ref{fig:f1}, which is referred to as Probe2. 
{The distribution is analysed via a bimodal fitting, with a parabola superimposed on a Gaussian function~\cite{Inada2008PRL,Schmiedmayer2010PRL}.}
From the bimodal fitting, we extract the coherent fraction so that the phase coherence of the dynamical many-body state can be inferred (see Supplementary Materials for more details).

\begin{figure*}
\begin{center}
\includegraphics[trim = 0mm 0mm 0mm 0mm, clip=true, width=.6\linewidth]{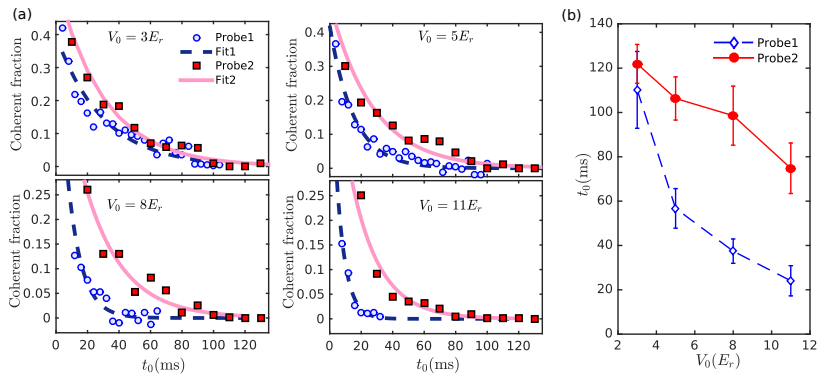}
\end{center}
\caption{ (Color online)  (a) The coherent fraction measured by Probe1,~2 with the different holding times, for $V_{0}=3 E_{r}$, $5 E_{r}$, $8 E_{r}$, and $11 E_{r}$, respectively. (b) The time $t_0$ for the atoms to lose coherence in lattice and pancake directions with different optical-lattice depths. The blue ``diamonds" are for $\hat{x}$-direction (lattice) by Probe1 and the ``dotted" points for the $\hat{y}$-direction (pancake). Errorbars are given by the $95\%$ confidence interval of fitting result. We find the time-dependence of the coherent fractions fits to a form of $Ae^{-t/\tau}$ with $A$ the amplitude and $\tau$ the characteristic time. The starting (ending) point of the sliding-phase superfluid is defined by  the time $t_0$ when the coherent fraction in the lattice (pancake) direction vanishes. With a relatively shallow lattice, say lattice depth $V_0 = 3 E_r$, there is essentially no difference in the time-dependence of the coherent fractions in lattice and pancake directions, which means a shallow lattice does not support the intermediate sliding-phase superfluid. With a deeper lattice, we find a significant difference in coherent fractions for lattice and pancake directions, leading to two dynamical timescales and an intermediate time window supporting the sliding-phase superfluid. The sliding-phase-lifetime can be systematically improved upon increasing lattice depth. Experimental results shown here are taken at temperature $120$~nK. } 
\label{fig:f3}
\end{figure*}

{\it Observation of the sliding-phase superfluid.---}
In order to characterize the real-time dynamics after P-band gets occupied, we measure momentum distributions in the lattice (Probe1) and pancake (Probe2) directions at the different holding-times  (Fig.~\ref{fig:f2}). Since the system is approximately rotation-symmetric in the $yz$-plane, the momentum distribution in the $\hat{z}$-direction is equivalent to that in the $\hat{y}$-direction, and is thus not shown here. From the time evolution, we identify three distinct dynamical regions. At early time---the first stage, the system has superfluid phase coherence in all three directions, which is clearly demonstrated through the sharp peaks observed in the momentum distribution shown in Fig.~\ref{fig:f2} at $t_0$ = 1 ms. A bimodal fitting shows the system is coherent in all directions. At late time---the final stage after about $100$~ms , the quantum gas has rethermalized with a complete loss of phase-coherence. The bimodal fitting (see Fig.~\ref{fig:f2}) shows all atoms are thermal in complete absence of any condensed component.   
There is yet an intermediate time region  with significant time-duration where the phase-coherence of the quantum system survives partially.  The bimodal fitting in Fig.~\ref{fig:f2} at $60$ ms shows that there is a finite condensed component in the pancake directions, but no such component in the lattice direction.   In this intermediate region the phase-coherence in the lattice direction already disappears, whereas the coherence in the pancake directions still persists as revealed by momentum distributions.

The evolution of phase-coherence in the three stages is described by a time-dependent correlation-function,
\begin{eqnarray}
&& \langle \hat{\phi} ^\dag  ({\bf r}, t)  \hat{\phi}({\bf r}', t)\rangle \nonumber \\
 & \propto &\exp (-|{\bf r}_x - {\bf r}'_x | /\xi_x  -|{\bf r}_y - {\bf r}'_y | /\xi_y  - |{\bf r}_z - {\bf r}'_z |/\xi_z),
\end{eqnarray}
where $\hat{\phi} ({\bf r}, t) $ is the bosonic field operator with spatial coordinate ${\bf r}$, $\xi_{x,y,z}$ the superfluid correlation length in  the three directions. 
At the first stage, the three correlation lengths diverge, or equivalently are comparable with the system-size, 
{whereas at the final stage the correlation lengths are all finite.} 
In the intermediate time region, we have divergent correlation lengths in the $yz$ plane, i.e.,  $\xi_{y,z} \sim L $ with $L$ the system-size, 
but finite correlation length in the $x$ direction,  i.e., $\xi_x/L \to 0$ .   
The peculiar dynamical phase in the intermediate time region represents the long sought-after sliding-phase superfluid---each pancake is phase coherent, but the relative phase across different pancakes is sliding. 
{We find that the sliding phase phenomenon is mainly supported by atoms in the P-band (see Supplementary Material).}

{\it Phase lifetime.---} To test the robustness of the dynamical sliding-phase superfluid, we measure its lifetime in dynamics. In the experiment, the lifetime is defined from time-dependent phase-coherent fractions in pancake and lattice directions, which are extracted from the momentum distributions (see Supplementary Materials). Experimental results are shown in Fig.~\ref{fig:f3}. We find that with the total atom number fixed in the experiment, there appears a critical lattice depth---$V_0=3E_r$ in our experiment---beyond which the sliding phase superfluid starts to emerge. 
{We expect the phase coherence starts to form from the pancakes in the trap center as those ones have relatively larger density and consequently larger superfluid stiffness. 
It is worth emphasizing that at  $V_0 = 3 E_r$  the P-band tunnelling is still significant (around $0.5 E_r$~\cite{Lireview2016}). }
As we increase the optical lattice depth further, the sliding phase superfluid becomes more robust in dynamics. For both lattice and pancake directions, a deeper lattice causes an overall decrease in the decay time of the coherence, owing to the increase of the interaction which accelerates the condensate depletion. But the coherence-decay-time for the lattice direction is affected more compared to the pancake directions, leading to a widening time-window that supports the intermediate sliding-phase superfluid. 

\begin{figure}
\begin{center}
\includegraphics[trim = 0mm 0mm 0mm 0mm, clip=true, width=.95\linewidth]{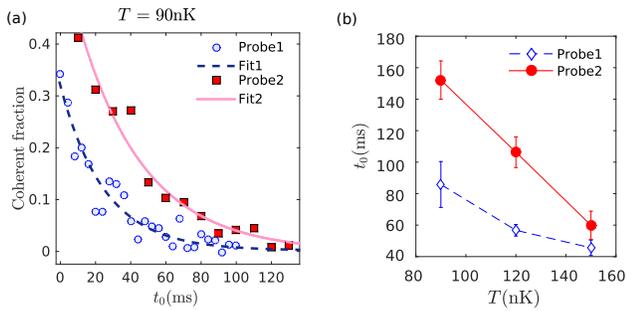}
\end{center}
\caption{ (Color online) (a) The coherent fraction with different holding times for a lower atomic temperature $T$=90 nK at $V_{0}=5 E_{r}$. 
 (b) The time ($t_0$) for atoms to lose coherence in lattice and pancake directions at different temperatures, which is extracted from an exponential fit of the coherent fraction (see Supplementary Materials). By increasing the temperature, we find the intermediate time window supporting the sliding-phase gets smaller with larger thermal fluctuations. The temperature dependence indicates the sliding-phase-lifetime can be further improved by cooling down to lower temperature. Errorbars are given by the $95\%$ confidence interval of fitting result.} 
\label{fig:f4}
\end{figure}

For completeness, we also examine finite temperature effects on the dynamical sliding-phase. 
{We note here that the temperature in the following refers to that of the atomic gas before loading into the lattice.} 
Comparing the results at temperature $T=90$~nK (Fig.~\ref{fig:f4}(a)) with $T=120$~nK (Fig.~\ref{fig:f3}(a)), the phase coherence gets more robust against decay at lower temperature as expected. The lifetime of the sliding phase superfluid depends on the relative coherence-robustness in lattice and pancake directions. 
{In the experiment, we find the coherence decay-time in the pancake direction is more prone to temperature effects compared to the lattice direction and increases more upon temperature decrease.  
This leads to a systematic increase in the lifetime of the intermediate  sliding phase superfluid (see Fig.~\ref{fig:f4}(b)) as the system is cooled down to a lower temperature.}  

{\it Phenomenological theory for the sliding phase.---} 
The emergence of the sliding phase in the time domain can be qualitatively captured by a $P$-band model~\cite{Lireview2016},  
\bea
 H &= &\textstyle \int d^2 {\bf r} \left \{ \sum_j p_j ^\dag ({\bf r}) \left[ -\frac{\hbar^2}{2M} \vec{\nabla}^2-\mu - V_{\rm trap} ({\bf r}) \right] p_j ({\bf r}) \right.  \nn \\
&+&   \left.  \textstyle J_p \sum_{<j,j'>}  {  p_j ^\dag ({\bf r}) p_{j'} ({\bf r}) } +  g \sum_j {p_j ^\dag p_j ^\dag p_j p_j}  \right\} , 
\eea
where $p_j$ ($p_j^\dag $) is the annihilation (creation) field operator, $<j,j'>$ represents nearest neighbouring lattice sites,  $V_{\rm trap}$ is the harmonic trap potential, $J_p$ is the tunnelling in the lattice direction, and $g$ represents the interaction strength. 
Atoms are initially prepared at the $P$-band maximum, so the system is dynamically unstable~\cite{Wu2001PRA,Martikainen2010,Xu2013}. 
The coherence in the lattice direction is then quickly lost, during which the kinetic energy in the lattice direction can be converted into kinetic energy in the pancake directions. This cross-dimensional energy transfer is expected to be the order of tunnelling $J_p$. Since the gas is continuous in each pancake, the in-plane ($yz$-plane) degrees of freedom would quickly relax and acquire an effective temperature description. The effective temperature of each pancake  is estimated from energy conservation to be 
\be 
k_B T_{\rm eff} \sim \left[N  (\hbar \omega)^2 J_p/L_x \right] ^{1/3}, 
\ee 
which is obtained at the weak interaction limit (see Supplementary Material). 
Here $k_B$ is the Boltzmann constant, and $\omega$ is the trap frequency in $yz$-plane, $N$ is the total particle number, $L_x$ the number of lattice sites in the $x$ direction. The number of thermal atoms is to the order of 
\be 
N _{\rm therm} \sim N \left[  (J_p/\hbar \omega)^2 L_x/N \right] ^{1/3} .
\ee  
The excessive atoms then remain condensed, giving rise to the phase coherence in each pancake. 
With particle number being fixed, we have a critical value of $J_p$, and consequently a critical lattice depth, for the sliding phase to emerge, which is qualitatively consistent with experimental observations. 
{In creating the sliding phase superfluid,  the role of the $P$-band is to enable an efficient preparation of the BEC in a dynamically unstable region using an adiabatic short-cut method developed in our experiments~\cite{load}.  Considering higher bands with odd parity is expected to support the dynamical sliding phase in a similar fashion.} 

On a microscopic level, modelling the sliding phase phenomena demands theoretical treatment of correlated dynamics in weakly-coupled XY-models beyond Gross-Pitaevskii treatment to carefully take into account thermal excitations. A quantitative description is expected to be nontrivial, for example, whether thermal atoms could act as effective-disorder previously proposed to stabilize the sliding-phase~\cite{Pekker} is worth consideration.

{\it Conclusion.---} 
To conclude, through quantum dynamics of ultracold atoms loaded in the excited band, our measurements unveil a sliding-phase superfluid. This sliding-phase appears due to thermalization time-scale separation for discrete and continuous degrees of freedom. The robustness of the dynamical phase has been tested by increasing lattice-depth and temperature.
This potentially opens up a novel route to search for metastable correlated phases with ultracold atoms driven far-out-of-equilibrium. The intricate exotic phases challenging to achieve in thermal-equilibrium or the quantum groundstate might find their natural realization in nonequilibrium settings. This work is also expected to shed light on understanding the high-Tc mystery in the nonequilibrium layered-Cuperates.

{\it Acknowledgement.---}
This work is supported by National Program on Key Basic Research Project of China (Grant No. 2016YFA0301501, Grant No. 2017YFA0304204), 
and National Natural Science Foundation of China (Grants No.11334001, and No. 61727819, No. 61475007, No. 91736208and No. 117740067).
XL also acknowledges support by the Thousand-Youth-Talent Program of China.








\begin{widetext} 

\newpage
\renewcommand{\theequation}{S\arabic{equation}}
\renewcommand{\thesection}{S-\arabic{section}}
\renewcommand{\thefigure}{S\arabic{figure}}
\renewcommand{\bibnumfmt}[1]{[S#1]}
\setcounter{equation}{0}
\setcounter{figure}{0}

\begin{center}
{\bf \Huge Supplementary Information}
\end{center}

\section{Shortcut method to load atoms into P-band}

{We have developed an effective method for transferring atoms from an harmonic trap into the desired band of an OL~\cite{load}. This shortcut is composed by standing-wave pulse sequences imposed on the system before the lattice is switched on. The time duration and interval in each step are optimized in order to reach the target state with a high fidelity and robustness. This process can be completed within several tens of microseconds. In this paper, the condensate is quickly loaded
into the P-band (first excited band) of the optical-lattice by two series of pulsed laser potentials.
In the first series of pulses the atom experiences spatial potential $V_{even}(x)=V_0\cos
^{2}\left( kx\right)$. In the second series, a shifted potential $V_{odd}(x)=V_0\cos^{2}\left(kx+\pi/4\right)$ is applied.
The relative spatial shift of the two pulses is introduced to efficiently change the parity from even to odd. The pulse sequence is designed to optimize the eventual P-band population.

In the experiment, two acousto-optic-modulators (AOM) with a frequency difference are used to form our
designed pulse sequence with the frequency difference
$\delta\omega=182.5$MHz which leads to a phase shift  of $5\pi/4$ between
two pulses.
We measure the P-band loading fidelity by the oscillations of the relative population in different momenta from the images by NAS~\cite{hu2015long}. By comparing the experimental data with the peripheral contour of the beating signal, the corresponding fidelity is found to be about 95\%.}

 \begin{figure}[htp]
\begin{center}
\includegraphics[trim = 0mm 0mm 0mm 0mm, clip=true, width=.55\linewidth]{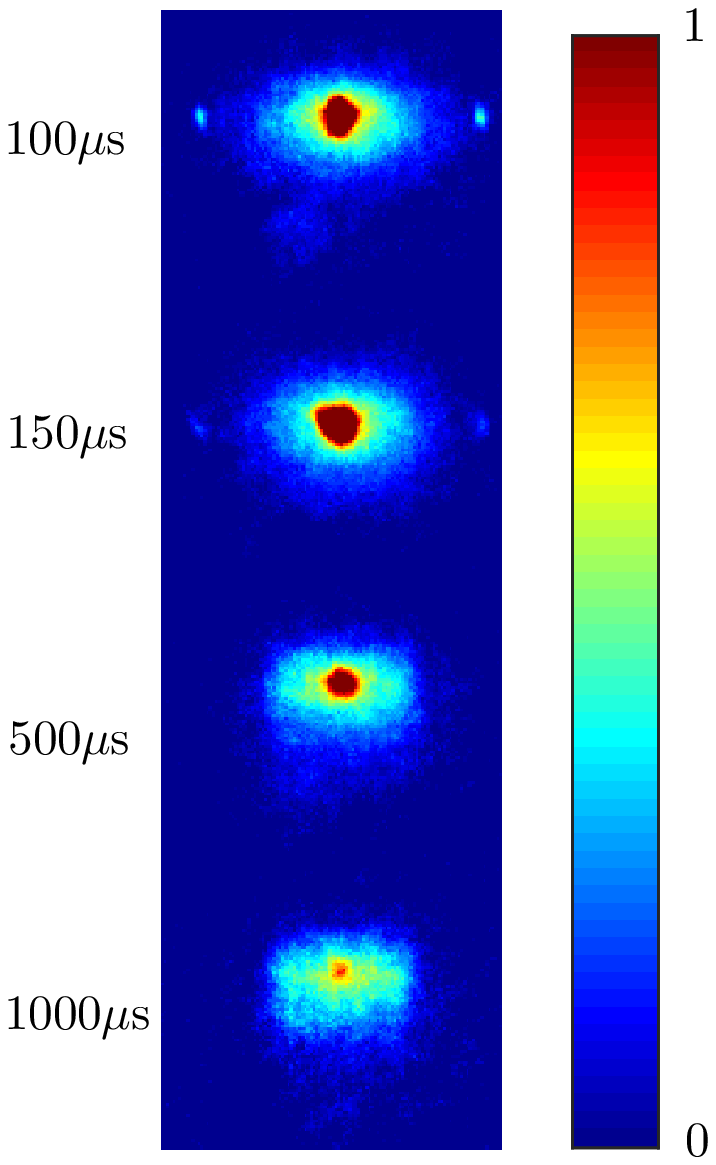}
\end{center}
\caption{(Color online) Results of the adiabatic ramping down of the lattice depth with different decay time. 
{The labeled time $100 \mu$s, $150 \mu$s, $500\mu$s, and $1000\mu$s represent the total time of lattice potential switch-off procedure in the band mapping.}  
}
\label{fig:S1}
\end{figure}

\section{Band mapping process}
In the band mapping process, it is required to have a proper choice over the band mapping time. It cannot be too short to ensure the inter-band dynamics must be adiabatic, and it cannot be too long so that the intra-band dynamics does not occur. To find such a proper time, we measure how the TOF results change with the total band mapping time.  The results are shown in Fig.~\ref{fig:S1}. The total time of band mapping process is varied from 100$\mu $s to 1000$\mu $s. 
We load the atoms to S-band with fidelity around $95\%$ to S-band of the 11$E_r$ lattice, and hold it in the lattice potential for 80 ms, then the atoms are released after a band map and detected with a TOF of 30 ms. The figure below shows how the probed atomic distribution would change with different band mapping time. For 100$\mu $s of total band mapping time, two side peaks at $\pm2\hbar k$  appear clearly, and likewise for 150$\mu $s. When the band mapping time reaches 500$\mu $s,   the atoms are converted to  the first Brilloiun zone, which implies the inter-band dynamics is already in the adiabatic limit.  Increasing the band mapping time further to 1000$\mu $s, there is no more inter-band atom transfer, but certain intra-band dynamics starts to occur. We thus choose a band mapping time of 500 $\mu$s in the experiments of mapping out atoms residing on  the S- and P-bands. We remark here that the choice of $500 \mu s$ refers to the total band mapping time. 
{The corresponding decay parameter $\eta$ in the ramp-down form of the potential---$V = V_0 e^{-t/\eta}$---shown in Fig.~1 of the main text is $100 \mu s$. }

\section{Finding coherent fraction}
To extract time dependence of coherent fractions in the lattice and pancake directions, we perform five experimental runs under each condition, and take the mean atomic distributions.
In the pancake directions (Probe2), the atomic distribution in the $\hat{y}$-direction is analyzed by a standard bimodal fitting according to
\begin{eqnarray}
&& f(y) = \\
&&G F_{th}(e^{-(y-y_0)^2/\sigma^2})
 + H(\max[1-(y-y_0)^2/\chi^2,0])^2, \nonumber
\end{eqnarray}
with the thermal distribution $F_{th}$ and a Thomas Fermi distribution for the condensate.
The parameters $G^\prime$ and $H$ are amplitudes of two components, $y_0$ is center of the atom cloud, and $\sigma$, $\chi$ represent widths of corresponding distributions. The coherent fraction is given by the atom number in the condensate component divided by the total atom number.
In the lattice direction (Probe1), we have coexistence of S- and P-band contributions in the atomic distributions which causes difficulty in extracting the coherent fraction in the system. In analyzing the measured atomic distribution, we take a general form of
\begin{eqnarray}
    f(x)= G^\prime f_T(x)  + f_C(x),
\end{eqnarray}
with $f_T(x)$ the thermal atom distribution, $f_C(x)$ the contribution from the condensed phase-coherent component.
The form of thermal-atom distribution $f_T(x)$ is fixed by using the atomic distribution in a parallel line with a $10$-pixel offset from the central line, where only thermal atoms exists.
The coherent component is then determined using the atomic distribution along the central line in the $\hat{x}$-direction (lattice direction).

\section{Determination of sliding-phase-lifetime}
In order to determine the lifetime of the sliding phase in the intermediate time-window, we need to find out the time $t_0$ where the coherent fractions in lattice and pancake directions vanish.
For both directions, we find that the decay of coherent fractions fit well to an exponential form $Ae^{-t/\tau}$ with different initial amplitude $A$ and decay time constant $\tau$. The vanishing time $t_0$ is determined when the coherent fraction $Ae^{-t/\tau}$ drops to its value of $1\%$, with the fraction $1\%$ the noise level in coherent  fraction measured for incoherent atoms.

\section{Absence of sliding phase with cold atoms in the $S$-band}

\begin{figure}[htp]
\begin{center}
\includegraphics[trim = 0mm 0mm 0mm 0mm, clip=true, width=.85\linewidth]{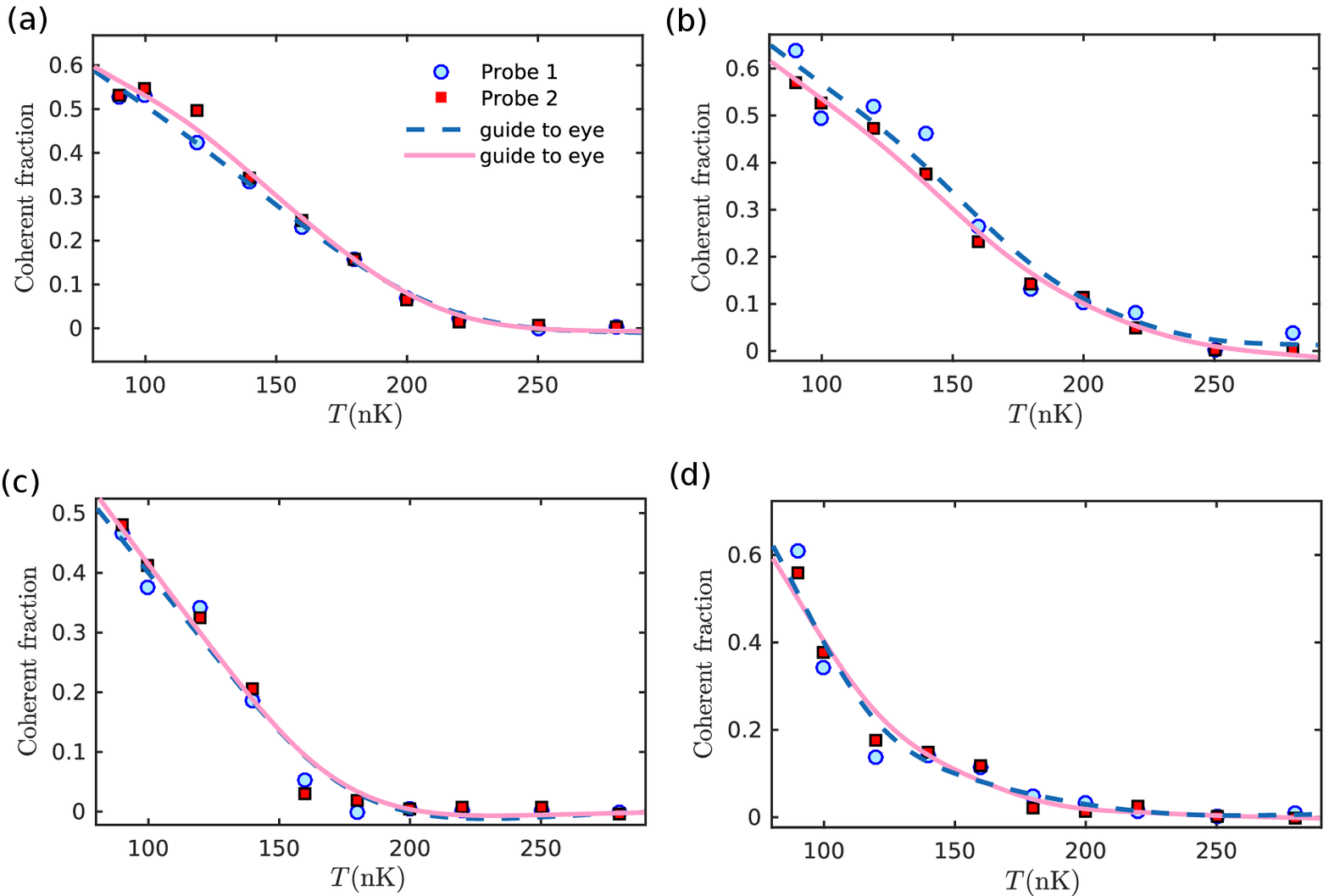}
\end{center}
\caption{(Color online) The coherence fraction in the lattice and axial directions for different lattice depths (a) 3$E_r$, (b) 5$E_r$, (c) 8$E_r$, (d) 11$E_r$ at different temperatures from 90 nK to 280 nK. The blue bashed line and the red solid line are smooth-curve-fitting of the result of Probe1 and Probe2, respectively.}
\label{fig:S2}
\end{figure}

In this section, we describe that P-band is necessary to realize the dynamical sliding phase in our experiment. To accomplish this, we carry out the experiments with atoms in the $S$-band only. 
The experiment begins with a pulse loading atoms to S-band, and the lattice potential is hold for 60 ms, then we do the same measurements as for the P-band. That is we probe the atomic distribution in $\hat{z}$-direction after a 30 ms TOF with a band mapping which is Probe1 in the main text, and also we measure the distribution by Probe2 in the $\hat{x}$-direction. The experiments are repeated 5 times for each parameter and then we take the mean distribution for the fitting. The coherence fraction we obtain from the axial and lattice directions would change with temperature. We measure the coherence fraction form 90 nK to 280 nK at four different lattice depth as shown in Fig.~\ref{fig:S2}.  The experimental results show that the coherence fraction would vanish in two directions simultaneously, which is distinct from the phase coherence pattern shown in Fig. 3 of the main text that corresponds to the sliding phase. These experimental observations are consistent with the fact that the Josephson coupled XY models are either in a completely incoherent normal phase or a three dimensional coherent superfluid phase, as concluded with field theory analysis~\cite{o1999sliding,Toner1990}. The sliding phase is thus absent for cold atoms  in the $S$-band at equilibrium.

\section{Coherence fraction in different bands}
\begin{figure}[t]
\begin{center}
\includegraphics[trim = 0mm 0mm 0mm 0mm, clip=true, width=.55\linewidth]{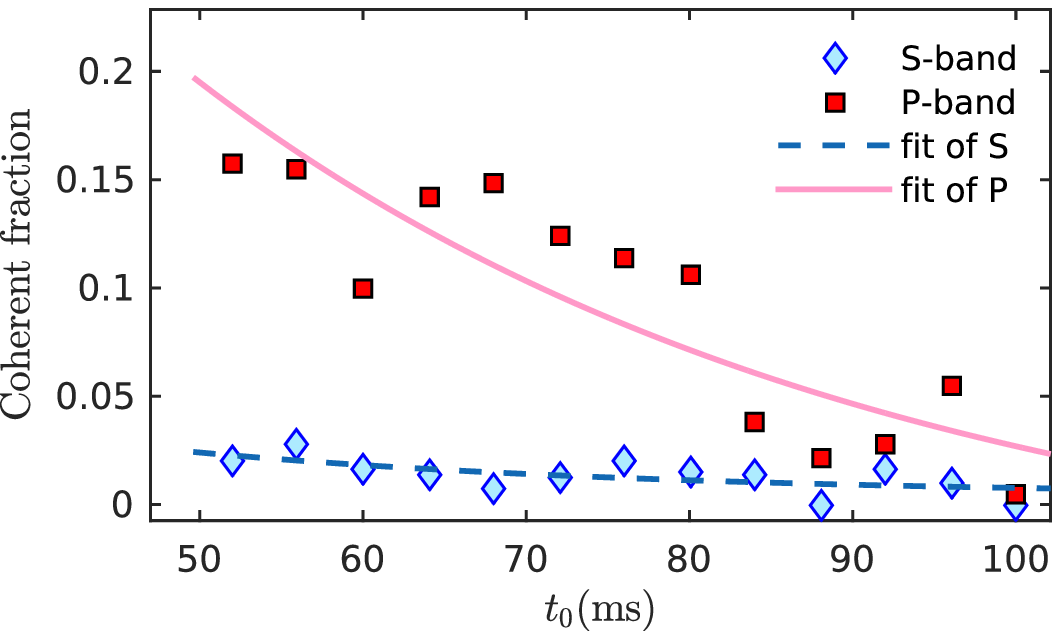}
\end{center}
\caption{(Color online) The coherence fraction in $\hat{y}$-directions fitted from S-band (blue diamonds) and P-band (red squares).  The lattice depth is $5E_r$ and the temperature of the atomic gas is $120$nK.}
\label{fig:S3}
\end{figure}

Since in our experiment we have finite atom population in both of the  S- and P-bands when the sliding phase occurs, it is crucial to characterize which band dominates the sliding phase phenomenon. Here we give our experimental results showing the P-band contribution is dominant for the sliding phase.  In the time window where the sliding phase exists, we can also fit the condensate distribution in y direction from the Probe1 image, which is with a larger noise compare to the case in Probe2. The distribution in y direction is calculated separately for region of S-band and P-band, as shown in the figure below. Before the fitting, we sum up the distribution in the first Brillouin zone for S-band, and the distribution in the second Brillouin zone for P-band. The result for lattice depth of $5E_r$ at 120 nK is shown in the Fig.~\ref{fig:S3}. 
In the time window where the phase coherence already disappears in the lattice direction, we find the coherent fraction in the P-band is significantly larger than that the S-band. For instance, at $t_0 = 60$ ms we get the condensate fraction in y direction as $f_S$  = 1.6$\%$ for S-band and $f_P  = 10.0\%$ P-band. 
We thus conclude that the sliding phase appears mainly in P-band.

\end{widetext}

\end{document}